\documentclass[aps,prb,floatfix,showpacs,twocolumn,superscriptaddress]{revtex4}
\usepackage{color}
\usepackage{bm}
\usepackage{amssymb}
\usepackage{amsmath}
\usepackage{amstext}
\usepackage{graphics}
\usepackage{epsfig}
\usepackage{placeins}
\usepackage{psfrag}
\usepackage{subfigure}

\begin{document}

\title{Superfluid-Mott-Insulator Transition in a One-Dimensional Optical
Lattice with Double-Well Potentials}
\author{H. C. Jiang}
\affiliation{Center for Advanced Study, Tsinghua University, Beijing,100084,China}
\author{Z. Y. Weng}
\affiliation{Center for Advanced Study, Tsinghua University, Beijing,100084,China}
\author{T. Xiang}
\affiliation{Institute of Theoretical Physics, Chinese Academy of
Sciences, P.O. Box 2735, Beijing 100080, China}
\date{\today}

\begin{abstract}
We study the superfluid-Mott-insulator transition of ultracold
bosonic atoms in a one-dimensional optical lattice with a
double-well confining trap using the density-matrix renormalization
group. At low density, the system behaves similarly as two separated
ones inside harmonic traps. At high density, however, interesting
features appear as the consequence of the quantum tunneling between
the two wells and the competition between the \textquotedblleft
superfluid\textquotedblright\ and Mott regions. They are
characterized by a rich step-plateau structure in the visibility and
the satellite peaks in the momentum distribution function as a
function of the on-site repulsion. These novel properties shed light
on the understanding of the phase coherence between two coupled
condensates and the off-diagonal correlations between the two wells.
\end{abstract}


\maketitle

\section{Introduction}

The experimental realization of trapped ultracold bosonic gases on
optical lattices has opened up the possibility of observing various
quantum phases (e.g., superfluid and Mott insulator) and studying
the quantum phase transitions between them in a well-controlled
manner. Without the confining trap, the system will undergo a
superfluid-Mott-insulator phase transition at integer fillings. In
the presence of a confining trap, the superfluid and Mott-insulator
regions may coexist\cite{Greiner02,Stoferle04} due to the
inhomogeneity induced by the trap.

Numerical calculations have been carried out for the systems
without{\cite{Till98,Till00}} and with a harmonic
trap\cite{Kashurnikov02,Batrouni02,Kollath04,Sengupta05}, using the
quantum Monte Carlo and density-matrix renormalization
group{\cite{white92,white93}} (DMRG) methods. Interesting new
features appear in the system with a single harmonic confining trap,
which are associated with the competition between the
\textquotedblleft superfluid\textquotedblright\ and Mott insulator
regimes and characterized by the \textquotedblleft
kinks\textquotedblright \cite{Sengupta05} in the visibility of
interference fringes with the increase of the on-site repulsion.

Recently a bosonic Josephson junction composed of two weakly coupled
Bose-Einstein condensates in a macroscopic double-well potential has
been experimentally realized{\cite{Albiez05}}. It raises an
interesting question concerning the quantum tunnelling effect in a
double-well potential system in which the ultracold atoms on optical
lattice are trapped. In such a system, the inter-atom interaction
and the competition between the \textquotedblleft
superfluidity\textquotedblright\ and Mott insulator phases are
expected to play a crucial role in shaping the inter-well
tunnelling.

In this paper, we intend to clarify this issue by studying a
one-dimensional Bose-Hubbard model in the presence of a double-well
potential using the DMRG. At low filling ($\rho =1/2)$, we find the
system simply behaves as two separated ones, each of which is in a
harmonic trap without significant coupling between them. At a higher
filling ($\rho =1$), however, the bosons in the two wells become
strongly correlated and a rich phenomenon related to the evolution
of the \textquotedblleft superfluid\textquotedblright\ and Mott
insulator regions as a function of the on-site repulsion is
obtained. A detailed discussion on how to detect these features
based on the visibility, momentum distribution, hopping correlation
function, and other physical quantities is given. These novel
properties shed light on the understanding of the phase coherence
and off-diagonal correlation between the coupled condensates in a
double-well interacting Bose system on optical lattices.

A cold atomic Bose gas on a one-dimensional optical lattice in the presence
of a double-well potential trap can be described by the following
Bose-Hubbard Hamiltonian
\begin{eqnarray}
H &=&-t{\sum_{i}}\left( \hat{a}_{i}^{\dagger }\hat{a}_{i+1}+h.c.\right)
\notag \\
&&+U{\sum_{i}}\hat{n}_{i}(\hat{n}_{i}-1)/2+{\sum_{i}}V_{T}(i)\hat{n}_{i},
\label{H}
\end{eqnarray}
where
\begin{equation}
V_{T}(i)=V_{T1}\left(i-\frac{L+1}{2}\right)^{2}+V_{T2}\left(i-\frac{L+1}{2}
\right)^{4}  \label{V}
\end{equation}
is the double-well trap with quadratic $V_{T1}$ and quartic $V_{T2}$
coefficients, and $L$ is the number of sites. The hopping integral
$t$ is set as the unit of energy $t=1$, and $U$ is the on-site
repulsion. $\hat{a} _{i}^{\dagger }$ and $\hat{a}_{i}$ are the
bosonic creation and annihilation operators, respectively, and
$\hat{n}_{i}=\hat{a}_{i}^{\dagger }\hat{a}_{i}$ is the number
operator.

An important parameter in characterizing the phase coherence in a
superfluid-Mott-insulator transition is the visibility of
interference fringes defined by {\cite {Gerbier05,Kollath05}}
\begin{equation}
\nu
={\frac{S_{\mathrm{max}}-S_{\mathrm{min}}}{S_{\mathrm{max}}+S_{\mathrm{
min}}}},  \label{nu}
\end{equation}
where $S_{\mathrm{max}}$ and $S_{\mathrm{min}}$ are the maximum and
minimum of the momentum distribution function
\begin{equation}
S(\mathbf{k})={\frac{1}{L}}{\sum_{i,j}}e^{i\mathbf{k}\cdot
(\mathbf{r}_{i}- \mathbf{r}_{j})}\left\langle \hat{a}_{i}^{\dagger
}\hat{a}_{j}\right\rangle .
\end{equation}
In the presence of a confining trap, it is useful to introduce the
local density of bosons $n_i=<\hat{n}_i>$ and the local particle
fluctuation {\cite{Batrouni02}}
\begin{equation}
\kappa _{i}\equiv \lbrack \left\langle \hat{n}_{i}^{2}\right\rangle
-n_{i}^{2}],  \label{kappa}
\end{equation}
to measure the inhomogeneity induced by the confining trap. To
further quantify the spatial correlations within the trap potential,
we have also calculated the hopping correlation function
\begin{equation}
g(i,j)=\left\langle \hat{a_{i}}^{\dagger }\hat{a_{j}}\right\rangle
\end{equation}
between two lattice sites $i$ and $j$, the total interaction energy
$E_U$,
\begin{equation}
E_{U} =U{\sum_{i}}n_{i}(n_{i}-1)/2,
\end{equation}
the kinetic energy $|E_k|$,
\begin{equation}
E_{k} =-t{\sum_{i}}\left\langle \hat{a}_{i}^{\dagger }\hat{a}
_{i+1}+h.c.\right\rangle ,
\end{equation}
and the trapping energy $E_T$
\begin{equation}
E_{T} ={\sum_{i}}V_{T}(i)n_{i},
\end{equation}
respectively.

\section{Superfluid-Mott-Insulator Transition}

In the following we present our DMRG calculations for the
Bose-Hubbard model (\ref{H}). Compared with a harmonic trap
\cite{Sengupta05}, the present system shows a richer and more
complex structure, due to the coupling between the two wells. When
the density is low enough such that there is no particle in the
middle of the system, the system splits into two weakly coupled
harmonic ones. However, if the density is sufficiently high that
particles in the two wells are not well separated, the interaction
between the wells will play a crucial role, which may change the
behavior of the system dramatically.

In the following, we will consider the system with 80 sites in a
double-well trap potential with $V_{T1}=-0.024t$ and
$V_{T2}=2\times10^{-5}t$. The calculations are carried out using the
DMRG with open boundary conditions, for two different filling
factors $\rho=1/2$ and $\rho=1$.

\subsection{Filling $\protect\rho=1/2$}

\begin{figure}[tbp]
\centerline{
    \includegraphics[height=3.0in,width=3.8in]{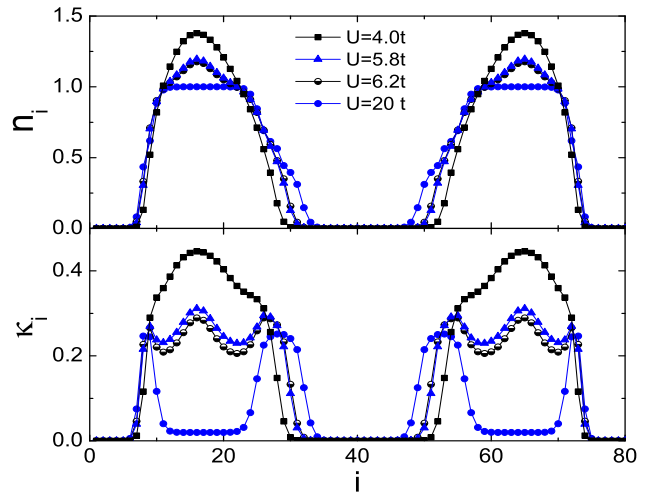}
    }
\caption{Density profiles $n_{i}$ and the corresponding particle fluctuation
$\protect\kappa _{i}$ for $\protect\rho =1/2$. The profiles for $U=5.8t$ and
$U=6.2t$ nearly coincide.}
\label{dw_den_kappa_p40s80}
\end{figure}

\begin{figure}[h]
\centerline{
    \scalebox{1.0}[1.0]{
    \includegraphics[height=4.0in,width=3.8in]{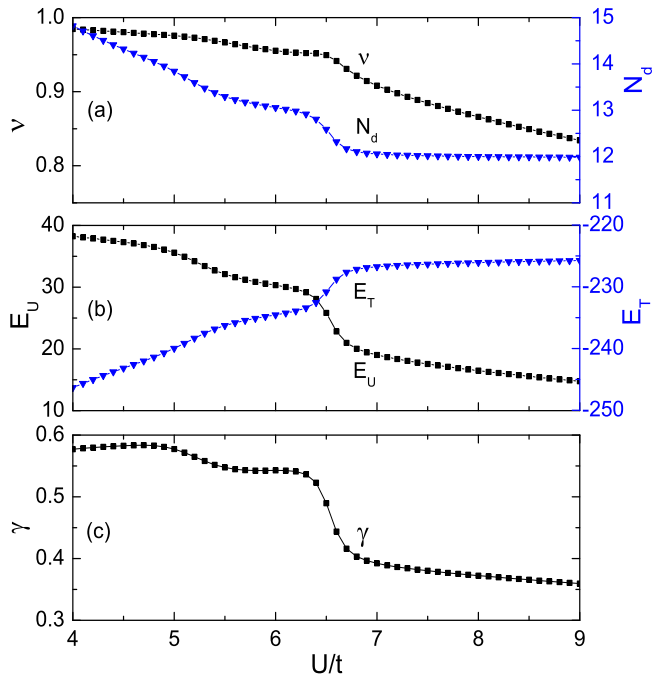}
    }}
\caption{(a) Visibility $\protect\nu $ and integrated density $N_{d}$; (b)
Interaction energy $E_{U}$ and trapping energy $E_{T}$; (c) Ratio $\protect%
\gamma =|E_{U}/E_{k}|$ of interaction to kinetic energy, as functions of $U/t
$, for the filling factor $\protect\rho =1/2$.}
\label{dw_visi_gamma_p40s80}
\end{figure}

AT the filling $\rho =1/2$, the system will not reach a
Mott-insulator state, even in the Tonks-Girardeau (TG) limit
$U\rightarrow \infty $, without a trap potential $V_{T}$. In the
presence of the double-well potential, however, two separated
Mott-insulator domains with the density $n_{i}=1$ appear inside the
two wells at large $U$, due to the low density of the system. Fig.
\ref{dw_den_kappa_p40s80} shows the density profile and
corresponding particle fluctuation at various values of $U/t$. At
small $U/t$, two separate superfluid regions appear in the two
wells. When $U$ is increased, the superfluid regions shrink until
the Mott-insulator domains emerge when $U/t$ is larger than $6.8$.

To further quantify this transition, let us define the integrated
density
\begin{equation}
N_{d}=\sum_{i=l_{1}}^{l_{2}}n_{i}  \label{nd}
\end{equation}
in a range $[l_{1},l_{2}]$. In the present system, we choose
$l_{1}=11$ and $ l_{2}=22$ in the left-side well. As shown in
Fig.\ref{dw_visi_gamma_p40s80} (a), $N_{d}$ is saturated to a
plateau of $N_{d}=12$ corresponding to a local Mott-insulator state
($n_{i}=1$) as $U/t$ is increased beyond $6.8$.

Fig.\ref{dw_visi_gamma_p40s80}(a) shows the visibility $\nu $. Two
kinks can be observed. The first one (less evident) occurs around $
U/t=5.5$, while the second one around $U/t=6.4$. As $U/t$ is
increased between $5.8$ and $6.2$, there is almost no change in the
density profile. This is due to presence of the emergent
Mott-insulator regions surrounding the superfluid ones in the two
wells. The atoms in the central superfluid regions can transfer to
the outer superfluid regions at larger $U/t$. This will eventually
exhaust the superfluid regions in the two well centers. At $U/t=20$,
the flat Mott-insulator plateaus, with vanishing local particle
fluctuation, centered around the two wells are clearly seen.

\begin{figure}[tbp]
\centerline{
    \includegraphics[height=2.2in,width=3.8in]{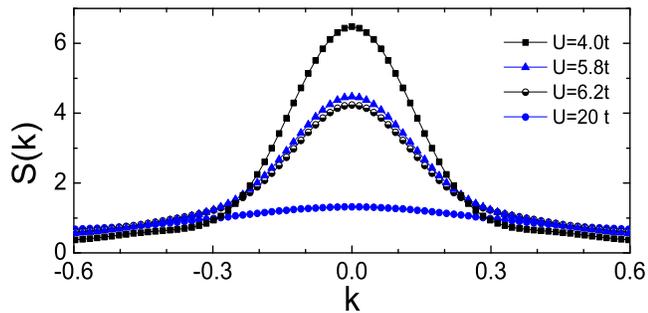}
    }
\caption{Momentum distribution function $S(k)$ for the system shown
in Fig. \protect\ref{dw_den_kappa_p40s80}.} \label{dw_sk_p40s80}
\end{figure}

As shown in Fig. \ref{dw_visi_gamma_p40s80}, the visibility with the
corresponding $E_{T}$, $E_{U}$ and the ratio $\gamma=|E_{U}/E_{k}|$,
exhibits a characteristic kink structure. The quick decrease of
$E_{U} $ in the interval $U/t=6.2-6.8$ shows a great suppression of
the double occupancy in the two wells. It is a manifestation of the
transition from superfluid to Mott-insulator in the two wells. This
occurs even though the total energy increases continuously with the
increase of $U/t$. By further increasing $U/t$, the density profile
remains almost unchanged.

Fig. \ref{dw_sk_p40s80} shows the corresponding momentum
distribution function $S(k)$ for the system shown in
Fig.\ref{dw_den_kappa_p40s80}. When $U/t$ is small, the system is in
a superfluid state and a single narrow peak appears at zero momentum
in $S(k)$.  For large $U/t$, $S(k)$ is broadened. The quantitative
change in $S(k)$ is best reflected in the kink structure of the
visibility $\nu $. In he Tonks-Girardeau limit, we find that
$S_{\mathrm{max}}=1.04$ and $\nu =0.44$. The non-zero visibility
$\nu $ reveals the presence of the superfluid regions surrounding
the Mott-insulator plateau regions (Fig. \ref{dw_den_kappa_p40s80}).

Finally, we note that at $\rho =1/2$, the distribution of bosons
remains well separated in the two wells
[Fig.\ref{dw_den_kappa_p40s80}]. Therefore, it can be approximately
treated as two single wells with weak coupling. Indeed, the results
presented above are consistent with those obtained in a harmonic
trap{\cite{Sengupta05}}.

\subsection{Filling $\rho =1$}

At $\rho=1$, bosons inside the two wells cannot be separated for
sufficiently large $U$. The system is in a Mott-insulator state
without the trap in the TG limit.

\begin{figure}[tbp]
\centerline{
    \includegraphics[height=4.0in,width=3.8in]{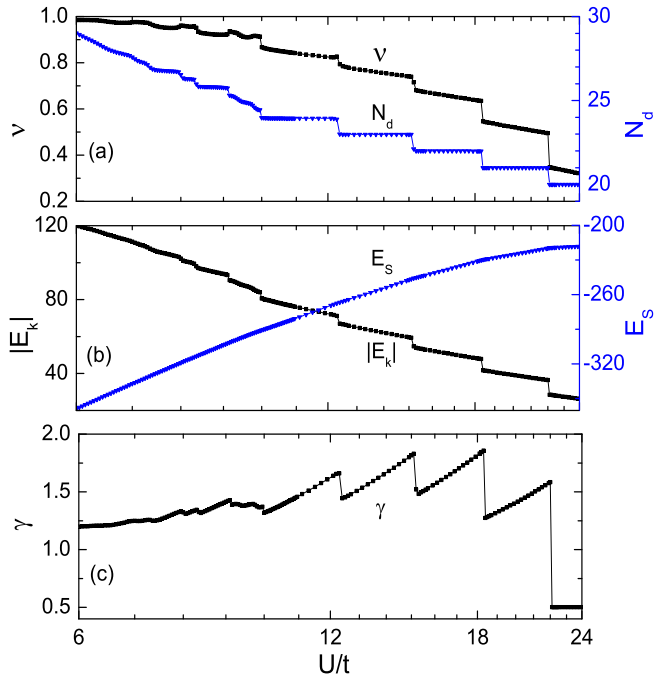}
    }
\caption{(a) Visibility $\protect\nu $ and integrated density $N_{d}$ from $%
l_{1}=11$ to $l_{2}=30$, (b) Kinetic $|E_{k}|$ and total energy $E_{S}$, (c)
Ratio $\protect\gamma =|E_{U}/E_{k}|$ of interaction to kinetic energy, as
functions of $U/t$ for $\protect\rho =1$.}
\label{dw_visi_gamma_p80s80}
\end{figure}

\begin{figure}[tbp]
\centerline{
    \includegraphics[height=2.6in,width=3.8in]{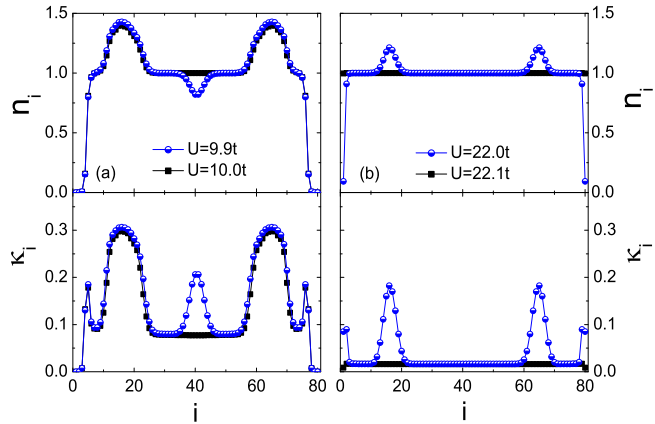}
    }
\caption{Density profiles $n_{i}$ and the corresponding local particle
fluctuation $\protect\kappa _{i}$ for $\protect\rho =1 $.}
\label{dw_den_kappa_p80s80}
\end{figure}

Fig.\ref{dw_visi_gamma_p80s80}(a) shows the visibility $\nu$ and the
integrated density $N_d$ ($l_1=11$, $l_2=30$) defined in
Eq.(\ref{nd}). The evolution of these two quantities exhibit a rich
step-plateau structure compared with the case of $\rho=1/2$. When
$U$ is small (about $\leq 10t$), the central Mott-insulator domain
has not formed. In this parameter regime, the plateaus in $\nu$ and
$N_d$ are quite narrow, indicating that the Mott-insulator domains
surrounding the superfluid ones are not robust and the bosons can
escape from the superfluid regions in the two wells. This feature is
well reflected in the density profiles and local particle
fluctuation around the step between $U/t=9.9$ and $10$ as shown in
Fig.\ref{dw_den_kappa_p80s80}(a).

When $U/t$ is further increased, the central region of the system
shows a flat Mott-insulator plateau with $n_{i}=1$ [Fig.
\ref{dw_den_kappa_p80s80}]. This leads to a sudden decrease in $\nu$
and $N_{d}$. Due to the large on-site repulsion $U$ and the
existence of the central Mott-insulator regions, bosons can only
tunnel from the inner superfluid region to the outer ones. Whenever
a boson escapes from the superfluid region, the visibility $\nu$ as
well as the integrated density $N_d$ shows a sharp decrease. When
$U/t$ is larger than $22$, the superfluid region disappears
competely (Fig.\ref{dw_den_kappa_p80s80}(b)). In the TG limit, the
whole system is in a Mott-insulator state and $\kappa _{i}=0$.

As shown in Fig. \ref{dw_visi_gamma_p80s80}(b), the abrupt changes
are also present in $ |E_{k}|$ at the steps of $\nu $, although the
total energy $E_{S} $ increases smoothly with $U/t$. The steps in
$|E_{k}|$ can be understood as the abrupt redistribution of the
boson density between the superfluid and Mott-insulator regions.
During this process, the superfluid region becomes smaller, while
the Mott-insulator becomes larger. The monotonic decrease of
$|E_{k}|$ with increasing $U/t$ is due to the widening of the Mott
insulator regions in which the kinetic energy is suppressed. In the
plateau regions, the increase in $\gamma$ is also due to the
reduction of the kinetic energy.

\begin{figure}[tbp]
\centerline{
    \includegraphics[height=3.8in,width=3.8in]{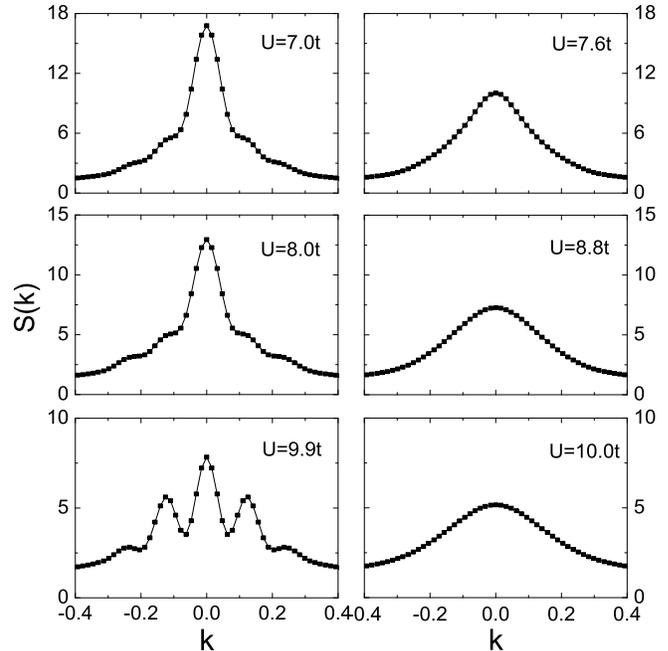}
    }
\caption{Momentum distribution function $S(k)$ at different values of $U/t$
for $\protect\rho =1$.}
\label{dw_sk_p80s80}
\end{figure}

Fig. \ref{dw_sk_p80s80} shows the momentum distribution function
$S(k)$. $S(k)$ exhibits different features at different $U/t$
regimes. Furthermore, $S(k)$ is correlated with $\nu $. When the
satellite peaks appear in $S(k)$, $\nu $ undergoes a sharp increase;
while $\nu $ undergoes a sharp decrease when the satellite peaks
disappear.

\begin{figure}[tbp]
\centerline{
    \scalebox{1.0}[1.0]{
    \includegraphics[height=3.8in,width=3.8in]{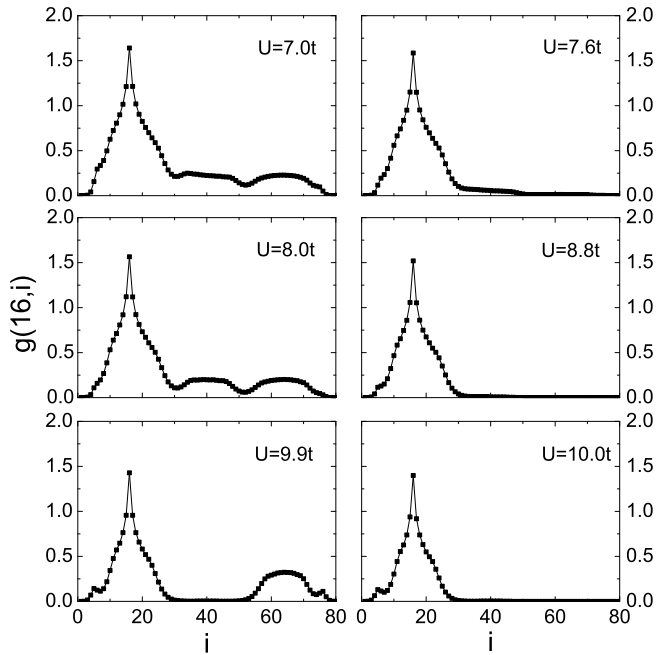}}
    }
\caption{Hopping correlation function $g(16,i)$ between site $16$ and $i$ at
different values of $U/t$ for $\protect\rho =1$.}
\label{dw_hop_p80s80}
\end{figure}

More detailed study demonstrates that the satellite peak structure
in $S(k)$ is related to the existence of a strong off-diagonal
long-range correlation between the two wells. This can be seen from
Fig. \ref{dw_hop_p80s80}, in which the hoping correlation function
$g(16,i)$ between site $16$ and site $i$ is plotted. If the system
has an off-diagonal long-range correlation, the satellite peaks are
present in $S(k)$. If the system does not have the off-diagonal
long-range correlation, the satellite peaks disappear. These
features are different from those in a harmonic
trap\cite{Kashurnikov02}. All the above features appear before the
formation of the central Mott-insulator plateau. In the presence of
the central Mott-insulator plateau, however, the satellite structure
of $S(k)$ as well as the long-range off-diagonal correlation between
the two wells disappears. This suggests that one can judge whether
or not the off-diagonal long-range correlation exists from the
measurement of momentum distribution functions.

\section{Summary}

To summarize, we have explored the superfluid-Mott-insulator
transition in a double-well trapped atomic Bose gas in a
one-dimensional optical lattice using the DMRG. We find that this
transition is well characterized by the visibility $\nu $, which
exhibits a series of kink structures associated with the
redistribution of bosons between local superfluid and Mott-insulator
regions, as a function of the on-site repulsion $U$. The evolution
of the system also shows a number of characteristic features in the
local integrated density, the density profile, local
compressibility, and the momentum distribution.

At low density (e.g. $\rho =1/2$), the system behaves like two
weakly coupled harmonic traps. With the increase of the density, the
system begins to develop novel properties unique to the double well
potential associated with a strong quantum tunneling effect between
the two wells. At high density (e.g. $ \rho =1$), a series of steps
and plateaus appear in the visibility and other physical quantities.
These features are related to the abrupt redistribution of the
bosons in the local superfluid and Mott-insulator regions. In
addition, the presence of the satellite peaks in the momentum
distribution function indicates that the strong off-diagonal
long-range correlation between the superfluid regions are separated
by the Mott insulator regions. Therefore, it is an effective probe
for experiment to identify the off-diagonal long-range correlation
and phase coherence due to the tunnelling between the two
condensates on the optical lattice.

\section{Acknowledgement}

We are grateful for helpful discussions with D.N. Sheng, F. Ye,
Y.C. Wen, X.L. Qi, Y. Cao, Z.C. Gu. Support from the NSFC grants
and the Basic Research Program of China is acknowledged.

\end{document}